\documentstyle[sprocl]{article}

\def\be{\begin{equation}}
\def\ee{\end{equation}}
\def\bea{\begin{eqnarray}}
\def\eea{\end{eqnarray}}

\def\today{\ifcase\month\orJanuary\or February\or 
March\or April\or May\or June\orJuly\or August\or 
September\or October\or November\or 
December\fi\space\number\day, \number\year}

\def\simlt{\lower.5ex\hbox{$\; \buildrel < \over \sim \;$}}
\def\simgt{\lower.5ex\hbox{$\; \buildrel > \over \sim \;$}}
\def\simpropto{\lower.2ex\hbox{$\; \buildrel \propto \over \sim \;$}}
\def\frac #1#2{{#1\over #2}}
\def\deg{\ifmmode^\circ\;\else$^\circ\;$\fi}
\def\solar{\ifmmode_{\mathord\odot}\;\else$_{\mathord\odot}\;$\fi}
\def\arcmin{\ifmmode^\prime\;\else$^\prime\;$\fi}
\def\arcsec{\ifmmode^{\prime\prime}\;\else$^{\prime\prime}\;$\fi}

\begin{document}

\title{SEVEN PARADIGMS IN STRUCTURE FORMATION}

\author{ Joseph Silk}

\address{Department of Physics, Astrophysics, 1 Keble Road,\\
Oxford OX1 3NP, England and
Departments of Astronomy and Physics, University of California, Berkeley, CA 94720}


\maketitle\abstracts{Have we converged on the definitive model of cosmology?
I present a critical assessment of the
current paradigms for the evolution of large-scale structure.}
\section{Introduction}

The current model for structure formation in the expanding universe has
been remarkably successful.  
Indeed it has recently been argued that we have
 resolved the principal issues in cosmology.\cite{turner}
However the lessons of history prescribe caution. 
There have been more oscillations
 in the values of the Hubble constant, the deceleration parameter and the
cosmological constant over the working life of a cosmologist than one cares
to recall. As the quality of the data has improved, one can be reasonably
confident that the uncertainties in parameter extraction have decreased.
But have we really converged on the definitive model? 

I have selected seven of the key paradigms
in order to provide a critical assessment.  To set the context I will
first review the reliability of the fundamental model of cosmology,
the Big Bang model, in terms of the time elapsed since the initial
singularity, or at least, the Planck epoch, $10^{-43}\rm s$.

Galaxies are well studied between the present epoch, 
$\sim 14\times 10^9 \rm yr,$
 and $\sim 3\times 10^9 \rm yr$ ($z\approx3$).  One can examine
the distribution of  Lyman alpha clouds, modelling
chemical evolution  from the gas phase metal abundances,\cite{pei} 
and find large numbers of  young, star-forming 
 galaxies back to about $2\times 10^9 \rm  yr$ ($z\approx 4$).\cite{steidel}
Beyond this are the dark ages where neither gas nor evidence of galaxy
formation has yet been detected.  Strong circumstantial evidence from
the Gunn-Peterson effect, indicating that the universe is highly
ionized by $z=5$, suggests that sources of ionizing photons must have
been present at an earlier epoch.  

Microwave background fluctuations
provide  substantial evidence on degree angular scales
for an acoustic peak,  generated at
$3\times 10^5 \rm yr $  ($z=1000$), when the radiation underwent its last
scatterings with matter.\cite{coble}  The blackbody spectrum of the cosmic
microwave background, with no deviation measured to a fraction of a
percent and a limit on the Compton  $y$  parameter
$\Delta y < 3 \times 10^{-6}  (95\% CL)$ 
on 7 degree angular scales,\cite{fixsen}
 could only have been generated in a sufficiently dense phase
which occurred during the first year of the expansion.  Light element
nucleosynthesis is an impressive prediction of the model, and testifies
to the Friedmann-like character at an epoch of one second. At this epoch,
neutrons first froze out of thermal equilibrium to subsequently become
incorporated in $^2$H, $^4$He, and $^7$Li, the primordial distribution
of which matches the predicted abundances for a unique value of the
baryon density.\cite{schramm}

Thus back to one second, there is strong observational evidence for the
canonical cosmology.  At earlier epochs, any observational predictions
are increasingly vague or non-existent.  One significant epoch is that
of
the quark-hadron phase transition ($t\sim 10^{-4} \rm s$ , $T\sim 100\rm\, MeV$),
which while  first order
cannot have been sufficiently inhomogeneous to amplify
density  fluctuations to form any primordial black holes.\cite{schmid}
The
electro-weak phase transition ($t\sim 10^{-10}\rm s$ , $T\sim 100\rm \, GeV$),
was even more short-lived but may have triggered baryon genesis.  Before then, one has the GUT
phase transition ($t\sim 10^{-35}\rm s$, $T\sim 10^{15}\rm  GeV$), and the
Planck epoch ($t\sim 10^{-43}\rm s,$ $T\sim 10^{19}\rm GeV$), of unification
of gravitation and electroweak and strong interactions.  Inflation is
generally believed to be associated with a strongly first order GUT phase
transition, but is a theory that is exceedingly difficult, if not
impossible, to verify.\cite{liddle}
  A gravitational radiation background at low
frequency is one possible direct relic of quantum gravity physics at the
Planck epoch, but we are far from being able to detect such a background.

In summary, we could say that our cherished beliefs, not to be
abandoned at any price, endorse the Big Bang model back to an
epoch of about one second
or $T\sim 1\rm \, MeV.$  One cannot attribute any comparable degree of
confidence to descriptions of earlier epochs because any fossils are
highly elusive.  Bearing this restriction in mind, we can now assess
the paradigms of structure formation.  The basic framework is provided
by the hypothesis that the universe is dominated by cold dark matter,
seeded by inflationary curvature fluctuations.  This does remarkably
well at accounting for many characteristics of large-scale structure in
the universe.  These include galaxy correlations on scales from 0.1 to
50 Mpc, the mass function and morphologies of galaxy clusters, galaxy
rotation curves and dark halos, the properties of the intergalactic
medium, and, most recently, the strong clustering found for luminous
star-forming galaxies at $z\sim 3$.  I will focus on specific paradigms
that underly these successes and assess the need for refinement both in
data and in theory that may be required  before we can be confident 
that we have found the ultimate model of cosmology.

\section{Paradigm 1:  Primordial Nucleosynthesis Prescribes
the Baryon Density}

Primordial nucleosynthesis predicts the abundances of several light
elements, notably $^2$H, $^4$He, and $^7$Li.  The principle variable is
the baryon density, $\Omega_{\rm b}\,h^2$.  One finds approximate
concordance for $\Omega_{\rm b}\,h^2\approx 0.015$, and with the
consensus value \cite{freedman} of H$_0$ ($h=0.7\,\pm 0.1$)  one concludes that
$\Omega_{\rm b} \approx 0.03$.  Not all pundits agree on concordance,
since the primordial $^4$He abundance requires a somewhat uncertain
extrapolation from the most metal-poor galaxies with He emission
lines ($Z\sim 0.02$ Z\solar) to zero metal abundances.\cite{olive} 
 Moreover the $^2$H abundance is
based on intergalactic (and protogalactic) $^2$H observed in absorption
at high redshifts toward  two quasars, probing only a very limited
region of space.\cite{burles}  However incorporation of $^7$Li and allowance for the
various uncertainties still leaves relatively impressive agreement
with simple model predictions.

Direct measurement of the baryon density at $z\sim 3$ can be
accomplished by using the Lyman alpha forest absorption systems toward
high redshift quasars.  The neutral gas observed is only a small
component of the total gas, but the ionizing radiation from quasars is
measured.  A reasonably robust conclusion finds that $\Omega_{\rm
	gas}\sim 0.04$, implying that the bulk of the baryons are observed and
in diffuse gas at high redshift.\cite{wein}

At low redshift, the luminous baryon component is well measured, and
amounts to $\Omega_*\sim 0.003$ in stars.\cite{fukugita}  Gas in rich clusters amounts
to a significant fraction of cluster mass and far more than the stellar
mass, but these clusters only account for about five percent of the
stellar component of the universe.  Combining both detected gas and
stars implies that at $z\sim 0$, we observe no more than $\Omega_{\rm
gas}\sim 0.005$.  Here we have a problem:  where are the baryons
today?

Most baryons must therefore be relatively dark at present.  There are
two possibilities, neither one of which is completely satisfactory.
The dark baryons could be hot gas at $T\sim 10^6$ K, in the
intergalactic medium.\cite{cen}
  This gas cannot populate galaxy halos, where it
is not observed, nor objects such as the Local Group, and is not
present in rich clusters in a globally significant amount.  It remains to be
detected:  if the temperature differed significantly from $10^6$ K the
presence of so much gas would already have had observable
consequences.

The alternative sink for dark baryons is in the form of compact
objects.  MACHOs are the obvious candidate, detected by gravitational
microlensing by objects in our halo of stars in the LMC, and possibly
 constituting
fifty percent of the dark mass of our halo.
However star-star lensing provides a possible alternative explanation of the
microlensing events,  associated with a previously undetected tidal
stream in front of the LMC \cite{lin}$^,$\cite{zhao} and  with the known extension
of the SMC along the line of sight. In the LMC case, at least one out of
approximately 20 events has a  known LMC distance, and for the SMC, there are only
two events, both of which are associated with the SMC.\cite{eros} The statistics are
unconvincing, and since until now one requires binary lenses to
 obtain a measure of the distance, any distance determinations are 
 likely to be biased towards star-star
lensing events.
\section{Paradigm 2: $\Omega = 1$}

It is tempting to believe that $\Omega_{\rm m}$ is unity.  If it is not
unity, one has to fine tune the initial curvature to one part in
$10^{30}$.  Moreover inflationary models generally predict that
$\Omega$ is unity.  However the evidence in favor of low $\Omega_{\rm
m}$, and specifically $\Omega_{\rm m}\approx 0.3$ is mounting.  The most
direct probe arises from counting rich galaxy clusters, both locally
and as a function of redshift.  The direct prediction of
$\Omega_{\rm m}=1$ is that there should be 
a higher-than-observed local density of clusters,
and strong evolution in number with redshift that is not seen.\cite{bah}
However this conclusion has recently been disputed.\cite{bla}$^,$\cite{via}

An indirect argument comes from studies of Type Ia supernovae, which
provide strong evidence for acceleration. This is most simply
interpreted in terms of a positive cosmological 
constant.\cite{perl}$^,$\cite{ries} The SN Ia data actually measure
$\Omega_{\Lambda}-\Omega_{\rm m}$.  Combined with direct measures of
$\Omega_{\rm m}$ both from galaxy peculiar velocities and from
clusters, one infers that $\Omega_{\Lambda}\approx 0.7$.  Hence
flatness is likely, and certainly well within observational
uncertainties.  Further evidence for the universe being spatially flat
comes from the measurement of the location of the first acoustic peak
in the cosmic microwave background anisotropy spectrum.  The location
reflects the angular size subtended by the horizon at last scattering,
and has Fourier harmonic $\ell=220\,\Omega^{-1/2}$.  Current data
requires \cite{web} $\Omega \simgt 0.4$, where $\Omega=\Omega_{\rm
m}+\Omega_{\Lambda}$.

Some possible pitfalls in this conclusion are that unbiased cluster surveys
have yet to be completed.  Use of wide field weak lensing maps will go a
long way towards obtaining a definitive rich cluster sample.  There is no
accepted theory for Type Ia supernovae, and it is possible that
evolutionary effects could conspire to produce a dimming that would mimic
the effects of acceleration, at least to $z\sim 1$.
Utilization of supernovae at $z>1$ will eventually help distinguish
evolutionary dimming or gray dust, the effects of
 which should be stronger at earlier epochs and hence 
with increasing $z$, from the
effect of acceleration, which decreases at earlier epochs,
that is with increasing $z$.

\section{Paradigm 3:  Density Fluctuations Originated in Inflation}

There is an elegant explanation for the origin of the density fluctuations
that seeded structure formation by gravitational instability.  Quantum
fluctuations are imprinted on a macroscopic scale with a nearly
scale-invariant spectral distribution of amplitudes, defined
by constant amplitude density fluctuations  at horizon
crossing.  This leads to a bottom-up formation sequence as the smallest
subhorizon scales acquire larger amplitudes and are the first to go
nonlinear.  One can compare the predicted linear fluctuations over scales
$\simgt 10$ Mpc with observations via microwave background fluctuations and
galaxy number count fluctuations.  $\delta T/T$ measures $\delta\rho/\rho$
at last scattering over scales from $\sim 100 $ Mpc up to the present
horizon.  Temperature fluctuations on smaller scales are progressively
damped by radiative diffusion, but a signal is detectable to an angular
scale of $\sim 10\arcmin$, equivalent to $\sim 20$ Mpc.  The conversion
from $\delta T/T$ to $\delta\rho/\rho$ is model-dependent, but can be
performed once the transfer function is specified.  At these high
redshifts, one is well within the linear regime, and if the fluctuations
are Gaussian, one can reconstruct the density fluctuation power spectrum.

Deep galaxy surveys yield galaxy number count fluctuations, which are
subject to an unknown  bias between luminous and dark matter.  Moreover, all
three dimensional 
surveys necessarily utilize redshift space.  Conversion from redshift space
to real space is straightforward if the peculiar velocity field is
specified.  One normally assumes spherical symmetry and radial motions on large
scales, and isotropic motions on scales where virialization has occurred,
with an appropriate transition between the linear  and nonlinear regimes.
On the virialization scale, collapse by of order a factor of 2 has occurred
in the absence of dissipation, and correction for density compression must
also be incorporated via interpolation or preferably via simulations.

Comparison of models with data is satisfactory only if the detailed
shape of the power spectrum is ignored.\cite{gawi}  A two parameter fit, via
normalisation at $8\,h^{-1}$ Mpc and a single shape parameter $\Gamma\equiv\Omega h$,
is often used.  For example, as defined below,
 $\sigma_8\equiv (\delta\rho/\rho)_{\rm
rms}/(\delta n_{\rm g}/n_{\rm g})_{\rm rms},$ 
as evaluated at $8\,h^{-1}$
Mpc, equals unity for unbiased dark matter.  COBE normalisation of
standard cold dark matter requires $\sigma_8\approx 1$ but the cluster
abundance requires $\sigma_8 \approx 0.6$.  The shape parameter $\Omega
h = 1$ for standard cold dark matter, but $\Omega h \approx 0.3$ is 
favoured for an
open universe.  One can fit a model to the data
with $\sigma_8\approx 0.6$ and
$\Omega h\approx 0.3$.  However detailed comparison of
models and observations reveals that there is no satisfactory fit to
the power spectrum shape for an acceptable class of models.  There is
an excess of large-scale power near 100 Mpc.  This is mostly manifested
in the APM galaxy and cluster surveys, but is also apparent in the Las
Campanas redshift survey.\cite{landy}
\section{Paradigm 4:  Galaxy Rotation Curves are Explained by Halos of
Cold Dark Matter}

Galaxy halos of cold dark matter acquire a universal density profile.\cite{nfw}
This yields a flat rotation curve over a substantial range of radius,
and gives an excellent fit to observational data on massive galaxy
rotation curves.  There is a central density cusp ($\propto 1/r$) which
in normal galaxies is embedded in a baryonic disk, the inner galaxy
being baryon-dominated.

Low surface brightness dwarf spiral galaxies provide a laboratory where
one can study dark matter at all radii:  even the central regions are
dark matter-dominated.  One finds that there is a soft, uniform density
dark matter core in these dwarf galaxies.\cite{primack}
 It is still controversial
whether the CDM theory can reproduce soft cores in dwarf galaxies:  at
least one group finds in high resolution simulations that the core
profiles are even steeper than $r^{-1}$, and have not converged.\cite{moore}

Disk sizes provide an even more stringent constraint on theoretical
models.  Indeed disk scale lengths  
cannot be explained.\cite{nfw}$^,$\cite{nav} The difficulty lies in the
fact that if angular momentum is
conserved as the baryons contract within the dark halos,
approximately the appropriate amount of angular
momentum is acquired by tidal torques between neighbouring density
flutuations to yield correct disk sizes 
However simulations fail to confirm this picture. In practice,
 cold dark matter
and the associated baryons are so clumpy that massive clumps fall into
the center via dynamical friction and angular momentum is
transferrd outwards. Disk torquing by dark matter clumps
also plays a role. The result is that the final baryonic disks 
are far too small.
  The resolution presumably lies 
in gas heating associated with injection of energy
 into the gas via supernovae once the first massive stars have formed.
\cite{efs}$^,$\cite{som}

\section{Paradigm 5:  Hierarchical Merging
 Accounts for the Luminosity Function
and the Tully-Fisher Relation}

Galaxies form by a succession of mergers of cold dark matter halos,
the baryons dissipating and forming a dense core.  Isolated infall
plausibly 
results in  disk formation. Disk merging concentrates the gas into a dense spheroid. 
 The transition from linear theory to
formation of self-gravitating clouds occurs at an overdensity of about
$\delta_{\rm crit}\approx 200$.  A simple {\it ansatz} due to Press
and Schechter yields the mass function of newly nonlinear objects
$$\frac{dN}{dM} \propto M^{-2} \exp \left[ -\delta^2_{\rm cr}/\langle
(\delta\rho/\rho)^2\,(M,\,t)\rangle \right]\,,$$
where $\delta^2 \equiv \langle(\delta\rho/\rho)^2 \,(M,\,t)\rangle$ is
the variance in the density fluctuations.  The variance at $8
h^{-1}\,{\rm Mpc}$, $\delta_8$, is given by
$$\delta_8=(R/8h^{-1} {\rm Mpc})^{-\frac{n+3}{2}}\,(1+z)^{-1}$$ where
$n\approx -1$ on cluster scales but $n\approx -2$ on galaxy scales,
and $M=10^{15}\,\Omega h^{-1}(R/8h^{-1}{\rm Mpc})^3\,{\rm M}\solar$.
Of course the luminosity function rather than the mass function is
actually observed.  We define $\sigma \equiv \delta/\delta_{\rm g}$, where
$\delta_{\rm g}$ is the variance in the galaxy counts.  On cluster scales,
one finds that $\sigma_8\approx 0.6\, (\pm 0.1)$ yields the observed
density of clusters if $\Omega=1$.  More generally, $\sigma_8$ scales
as $\Omega^{-0.6}$.  A larger $\sigma$ is required for a given number
density of objects in order to account for the reduced growth in
$\delta$ as $\Omega$ is decreased below unity.

To match the observed luminosity function and predicted mass function
requires specification both of $\sigma_8$ and of the mass-to-light
ratio.  Much of the dark mass is in objects that were the first to go
nonlinear, as well as in the objects presently going nonlinear.  Hence
one crudely expects  that $M/L\approx 400\,h$,
as measured in rich clusters. The global value of $M/L$ is
$M/L\approx 1500\Omega h,$ and happens to coincide
with the mass-to-luminosity ratio measured for rich clusters if $\Omega
\approx 0.4.$ This suggests that these clusters may provide a fair sample of the universe.
Even if most dwarfs do
not survive, because of subsequent merging, the relic dwarfs are
expected to have high $M/L$.  Later generations of galaxies should have
undergone segregation of baryons, because of dissipation, and the
resulting $M/L$ is reduced.  Many of the first dwarfs are disrupted to
form the halos of massive galaxies.  The predicted high $M/L$ (of order
100) is consistent with observations, both of galaxy halos and of the
lowest mass dwarfs (to within a factor of $\sim 2$).

However it is the detailed measurement of $M/L$ that leads to a 
possible problem.
One has to normalise $M/L$ by specifying the mass-to-light ratio of
luminous galaxies.
The observed luminosity function can be written as
$$\frac{dN}{dL}\propto L^{-\alpha}\exp (-L/L_*)$$
where $\alpha\approx$ 1 -- 1.5, depending on the selection criterion, and
$L_* \approx 10^{10}\,h^{-2} {\rm L}\solar$.  Matching to the predicted
mass function specifies $M/L$ for $L_*$ galaxies, as well as the slope
of the luminosity function.  One forces a fit to $\alpha$ by invoking
star formation-induced feedback and baryonic loss.  This
preferentially reduces the number of low mass galaxies.  A typical
prescription is\,\cite{lacey} that the retained baryonic  fraction is given by
$$f_{\rm B}=(v_c/v_*)^2\, ,$$
where $v_c$ is the disk circular velocity.
Dwarfs are preferentially disrupted by winds.  In this way one can fit
$\alpha$.  There is no longer any freedom in the luminous galaxy
parameters.

Potential difficulties arise as follows.  Simulations of mass loss from
dwarf galaxies suggest that supernova ejecta may contribute to
the wind
but leave much of the interstellar gas bound to the galaxies.
\cite{mac}  This
would be a serious problem as one relies on redistribution of the
baryonic reservoir to form massive galaxies.  Another problem arises
with the Tully-Fisher relation.  This is the measured relation,
approximately $L\simpropto V^{\beta}_{\rm rot},$ between galaxy
luminosity and maximum rotational velocity.  In effect, the
Tully-Fisher relation offers the prescription
for $M/L$ within the luminous part of the
galaxy, since the virial theorem requires
$$L\approx V^4_{\rm rot}\,G^{-2}\,\mu^{-1}_L\,(L/M)^2$$
where $\mu_L$ is the surface brightness of the galaxy. Since $\mu_L$
has a narrow dispersion for most disk galaxies, the Tully-Fisher
relation, where $\beta\approx 3$ is measured in the  $I$ band and 
$\beta\approx 4$
is appropriate to the near infrared, effectively constrains $M/L$.  The
normalization of the Tully-Fisher relation requires $M/L\approx 5h$
for early-type spirals, as is observed directly from their rotation
curves within their half-light radii.  However simulations of
hierarchical clustering, which incorporate baryonic cooling and star
formation with a prescription designed to reproduce the luminosity
function, give too high a normalization for $M/L$ in the predicted
Tully-Fisher relation:  at a given luminosity the rotational velocity
is too high.\cite{navs}
  Moreover the efficient early star formation required in
order to fit the luminosity function requires the Tully-Fisher
normalisation to change with redshift:  galaxies are 
predicted to be brighter by about
a magnitude at a given rotation velocity at  $z \sim 1$,
and this exceeds the observed offset.\cite{vogt}  Resolution of
the Tully-Fisher normalization remains controversial.

\section{Paradigm 6:  The Bulk of the Stars Formed After $z=2$}

Identification of the Lyman break galaxies, by using the 912 \AA\ 
discontinuity in predicted spectra as a broad band redshift indicator,
has revolutionized our knowledge of early star formation.  Current
samples of high redshift star-forming galaxies, chosen in a relatively
unbiased manner, contain $\sim 1000$ galaxies at $z\sim 3$ and $\sim
100$ galaxies at $z\sim 4$.  The volume of the universe
involved is known, and one can therefore compute the comoving
luminosity density.\cite{stei} Since the galaxies are selected in the 
rest-frame
UV, one can convert luminosity density to massive star formation rate.
One uncertainty is correction for dust extinction but this is mostly
resolved by measurement of the galaxy spectra.

If, say, a Miller-Scalo  initial stellar 
mass function is adopted, one concludes that the star
formation rate per unit volume rose rapidly between the present epoch and
redshift unity by a factor of about 10.  Beyond redshift one, the star formation rate remains
approximately constant, to $z>4$.  Moreover the median star formation
rate per galaxy is high, around 30 M\solar per year, the star forming
galaxies are mostly compact, and strong clustering is found.\cite{ste}
One
interpretation of the data is that most stars formed late, because of the
short cosmic time available at high redshift, and that most of the
Lyman-break galaxies are massive, and hence clustered, objects that are
probably undergoing spheroid formation.  An alternative view is that
the clustering is due to merger-induced starbursts of low mass galaxies
within massive galaxy halos.\cite{pri} Reconcilation of either interpretation
with hierarchical clustering theory requires a low $\Omega$ universe,
especially in the former case, and a detailed prescription for galaxy
star formation.  The rapid rise in the number of
star-forming galaxies at low
redshift is especially challenging if $\Omega$ is low, since galaxy
clustering reveals little or no evolution at $z\simlt 1$, as
measured by cluster abundances, and both massive disk sizes and the
Tully-Fisher relation show little change to $z\sim 1$. 
One intersting suggestion is  that a new population of blue compact, star-forming galaxies is responsible for the evolution in the  star formation rate density of the universe.\cite{guz}
\section{Paradigm 7:  Galaxy Spheroids Formed Via Mergers}

Galaxy mergers are recognized as the triggers of nearby starbursts,
especially the ultraluminous far infrared-selected galaxies.  These
systems are powered in large part 
by star formation rather than by an embedded AGN,
as confirmed by far infrared spectroscopy, and have star formation
rates of 100 or even 1,000 M\solar per year.  Near infrared mapping
reveals de Vaucouleurs profiles and CO mapping reveals a central cold
disk or ring with $\sim 10^{10}$ M\solar of molecular gas within a few
hundred parsecs.  Can one generalize from the rare nearby examples that
ellipticals, and more generally spheroids, formed via merger-induced
starbursts?

Evidence that gives support to this contention requires a component of
star-forming galaxies that is sparse locally to account for three
distinct observations of galaxies, or of their emission presumed to be
at $z>1$.  Far IR counts by ISO at 175 $\mu$m \cite{elb}
and submillimeter counts
by SCUBA \cite{lilb} at 850 $\mu$m require a population of IR-emitting objects that
have starburst rather than normal disk infrared spectra.  Moreover
identification of SCUBA objects demonstrates that typical redshifts are
one or larger, but  mostly below 2.\cite{lil} 

A powerful indirect argument has emerged from 
modelling of the
diffuse far infrared background 
radiation.  This amounts to \cite{lagache} $\nu i_\nu \sim 20$ nw/m$^2$sr, and
exceeds the diffuse optical background light of about 10 nw/m$^2$sr that
is inferred from deep HST counts.  The local population of galaxies,
evolved backwards in time fails to account for the diffuse infrared
light, if one only considers disk galaxies, where the star formation
history is known from considerations of their dynamical evolution.

  The
starburst population invoked to account for the FIR counts can account
for the diffuse infrared background radiation.\cite{gui}
If this is the case, one
expects a non-negligible contribution near 1 mm wavelength from
ultraluminous FIR galaxies to the diffuse background radiation.  For
example, the predicted FIR flux peaks at $\sim 400\,\mu$m if the
mergers occur at $z\simlt 3$.  The extrapolation to longer wavelengths
tracks the emissivity, or decreases roughly 
as $\lambda^3$.  Hence there should be a
contribution at 1 mm of order 1 nw/m$^2$sr, which may be compared with
the CMB flux of $\sim 2000$ nw/m$^2$sr.  One can measure fluctuations of
$\delta T/T \sim 10^{-6}$, and one could therefore be sensitive to a
population of $\sim 10^6$ ultraluminous FIR sources at high $z$.  The
inferred surface density ($\sim 20$ per square degree) is comparable to
the level of current SCUBA detections.  Hence CMB fluctuations on an
angular scale of $\sim 10$\arcmin near the CMB peak could be generated
by the sources responsible for the diffuse FIR background.  Moreover
these are rare and massive galaxies, and hence are expected to have a
large correlation length that should give an imprint on degree scales.

One can evidently reconcile submillimeter counts, the cosmic star
formation history and the far infrared background together with
formation of disks and spheroids provided that a substantial part of
spheroid formation is dust shrouded.  A difficulty that arises is the
following: where are the precursors of current epoch 
 ellipticals? A few are seen
at $z<5$ but are too sparse in number to account for the younger counterparts
 of
local ellipticals.\cite{zepf}  Dust shrouding until after the A stars have faded
($\sim 2\times 10^9$ yr) would help. Other options are that the young
ellipticals are indeed present but disguised via ongoing star
formation activity, and mostly form at $z>5$ or else possess an IMF deficient
in massive stars.

\section{Conclusions}

 Cosmological model-building has made impressive advances in the
past year. However much of this rests on supernovae being standard
candles.  This is a demanding requirement, given that we lack complete
models for supernovae. Consider a Type I supernova, for which one
popular model consists of a close pair of white dwarfs.  We do not
know a priori whether a pair of merging white dwarfs will explode or
not, or will self destruct or leave a neutron star relic.  Other
models involve mass transfer onto a white dwarf by an evolving close
companion: again, we do not know the outcome, whether the endpoint is
violently explosive or mildly quiescent. No doubt some subset of
accreting or merging white dwarfs are SNIa, but we do not know how to
select this subset, nor how evolution of the parent system would
affect the outcome in the early universe.

One of the largest uncertainties in interpreting the SCUBA
submillimeter sources is the possible role of AGN and quasars in
powering the high infrared luminosities. The absence of a hot dust
component in some high redshift ultraluminous infrared galaxies
(ULIRGs) with CO detections argues for a star formation interpretation
of infrared luminosities as high as $10^{13}\rm L_\odot.$ Observations
of far infrared line diagnostics suggest thatup to  $\sim 20\%$ of ULIRGs
may be AGN-powered,\cite{gen}  but nearby examples such as Arp 220 suggest that
even in these cases there may be comparable amounts of star
formation-induced infrared luminosity. Interpretation of the hard
($\sim 30$ keV) x-ray background requires the mostly resolved sources
responsible for the background to be self-absorbed AGN surrounded by
dusty gas that reemits the absorbed AGN power at far infrared
wavelenghts and can at most account for $\sim 10-20 \%$ of the diffuse
far infrared background.\cite{alm} An independent
argument is as follows:
the correlation of central black holes in nearby
galaxies with spheroids ($M_{bh}\approx 0.005 M_\ast$)
suggests that with an accretion efficiency that is expected to be
a factor $f\sim 10-30$ larger than the nuclear burning efficiency for producing
infrared emission, the resulting contribution from AGN and quasars to the 
far infrared background should be $\sim 15 (f/0.03)\%$ of the contribution from star formation.

There are too many unresolved issues in the context of structure
formation to be confident that we have converged on the correct
prescription for primordial fluctuations in density, nonlinear growth,
and cosmological model.  And then we must add in the complexities of
star formation, poorly understood in the solar neighbourhood, let
alone in ultraluminous galaxies at high redshift.  One cannot expect
the advent of more powerful computers to simply resolve the
outstanding problems.  Rather it is a matter of coming to grips with
improved physical modelling of star-forming galaxies. Phenomenological
model building is likely to provide more fruitful returns than brute
force simulations, but the data requirements are demanding even on the
new generations of very large telescopes.  Fluctuation spectra will be
measured with various CMB experiments, although disentangling the
various parameters of cosmology and structure formation will take
time.

However I am optimistic that the anticipated influx of new data, from optical, infrared, x-ray  and radio telescopes will go far towards resolving these uncertainties. 
It is simply that the  journey will be long with many detours, before we  have deciphered the ultimate
model of cosmology.
\section*{Acknowledgments}
I thank Ana Mourao and Pedro Ferreira for the gracious hospitality provided
in Faro.


\section*{References}
\begin{thebibliography}{99}
\bibitem{turner} Turner, M., astro-ph/9811447 (1998), RMP (in press).
\bibitem{pei} Pei, Y., Fall, S. and Hauser, M. Moore,
astro-ph/9812182 (1998), ApJ in press.
\bibitem{steidel} Steidel, C. et al. astro-ph/9811400 (1998), to 
appear in the proceedings of the Xth Rencontres de Blois, "The Birth of
Galaxies", July 1998.
\bibitem{coble} Coble, K. et al. astro-ph/9902195 (1999), in press.
\bibitem{fixsen} Fixsen, D.  et al. ApJ {486}, {623} (1997).
\bibitem{schramm} Schramm, D. and Turner, M. RMP {70}, {303} (1998).
\bibitem{schmid} Widerin, P. and Schmid, C., preprint astro-ph/9808142
(1998).
\bibitem{liddle} Liddle, A. preprint astro-ph/9901124 (1999).
\bibitem{freedman} Freedman, W. {\it et al.} 1998, in IAU Symposium 183, {\it
 Cosmological Parameters and the Evolution of the Universe}, in press.
\bibitem{olive} Olive, K., Steigman, G. and Stillman, E. 1997, ApJ, 483,788
(1997).
\bibitem{burles} Burles, S. and Tytler, S.  ApJ {507}, {732} (1998).
\bibitem{wein} Weinberg, D. et al. ApJ, 490, 564 (1997).
\bibitem{fukugita} Fukugita, M., Hogan, C. and Peebles, P. ApJ {503}, {518}
 (1998)
\bibitem{cen} Cen, R. and Ostriker, J. ApJ, in press (1999).
\bibitem {lin} Zaritsky, D. and Lin, D. AJ, 114, 2545 (1997).
\bibitem {zhao} Zhao, H. MNRAS, 294, 139 (1998).
\bibitem{eros} Alard, C. et al. A\&A, 337, L17 (1998).
\bibitem {bah} Bahcall, N. and Fan, X., ApJ, 504, 1 (1998).
\bibitem {bla} Blanchard, A. et al., preprint astro-ph/9810318
(1998).
\bibitem {via} Viana, P. and Liddle, A., preprint astro-ph/9803244
(1998).
\bibitem {perl} Perlmutter, S. et al., preprint astro-ph/9812133, ApJ in press (1999). 
\bibitem {ries} Riess, A. et al.,  preprint astro-ph/9805201, AJ in press (1999). 
\bibitem{web} Webster, M. et al., preprint astro-ph/9802109
(1998).
\bibitem{gawi} Gawiser, E. and Silk, J. Science, 280, 140 (1998).
\bibitem{primack} Kravtsov, A.V. et al., ApJ {502}, {48} (1998).
\bibitem{landy} Landy, S. et al. ApJ {456}, {L1} (1996).
\bibitem{nfw}  Navarro,  J., Frenk, C.  and White, S. ApJ, {490}, 
{493}, {1997}.
\bibitem{moore} Moore, B. et al., ApJ {499}, {L5} (1998).
\bibitem{nfw}  Navarro, J., Frenk, C. and  White, S. MNRAS,  275, 56 (1995).
\bibitem{nav} Navarro, J. F. and Steinmetz, M., ApJ {502}, {48} (1998).
\bibitem{efs} Weil, M., Eke, V. and Efstathiou, G.,  MNRAS
{300}, {703} (1998).
\bibitem{som} Sommer-Larsen, J., Gelato, S. and Vedel, H., preprint astro-ph/9801094 (1998).
\bibitem {lacey}  Lacey, C. et al.,  ApJ, 402, 1 (1993).
\bibitem{mac} Maclow, M. and Ferrara, A., preprint astro-ph/9801237 (1998).
\bibitem{navs} Navarro, J. F. and Steinmetz, M., preprint astro-ph/9808076 (1998).
\bibitem{vogt} Vogt, N. et al. ApJ, 479, 121 (1997).
\bibitem{stei} Steidel, C. et al., preprint astro-ph/9811399 (1998).
\bibitem{ste} Steidel, C., preprint astro-ph/9811400 (1998).
\bibitem{pri} Somerville, R.,  Primack, J. and Faber, S.,
 preprint astro-ph/9806228 (1998).
\bibitem {guz}  Guzman, R. et al., ApJ, 489, 559 (1997).
\bibitem{lagache} Lagache, G. et al., preprint astro-ph/9901059 (1999).
\bibitem{elb} Elbaz, D. et al., preprint astro-ph/99022291999).
\bibitem{lilb} Lilly, S. et al., preprint astro-ph/9903157(1999).
\bibitem{lil} Lilly, S. et al., preprint astro-ph/9901047 (1999).
\bibitem{gui} Guiderdoni, B.  et al., preprint astro-ph/9902141 (1999).
\bibitem{zepf} Zepf, S., Nature, 390, 377 (1977).
\bibitem {gen}  Genzel, R. et al., ApJ, 498, 579 (1998).
\bibitem{alm} Almeini, O., Lawrence, A. and Boyle, B.,
 preprint astro-ph/9903178, MNRAS in press (1999)
\end{thebibliography}
\end{document}